\documentclass[preprint,showpacs,preprintnumbers,amsmath,amssymb]{revtex4}
\usepackage{graphicx,epsfig,dcolumn,bm,epic,eepic,float}% Include figure files
\usepackage{amsmath}
\usepackage{latexsym}
\usepackage{color}
\usepackage{makeidx,shortvrb,latexsym}
\begin{document}
\unitlength 1 cm
\newcommand{\be}{\begin{equation}}
\newcommand{\ee}{\end{equation}}
\newcommand{\bearr}{\begin{eqnarray}}
\newcommand{\eearr}{\end{eqnarray}}
\newcommand{\nn}{\nonumber}
\newcommand{\vk}{\vec k}
\newcommand{\vp}{\vec p}
\newcommand{\vq}{\vec q}
\newcommand{\vkp}{\vec {k'}}
\newcommand{\vpp}{\vec {p'}}
\newcommand{\vqp}{\vec {q'}}
\newcommand{\bk}{{\bf k}}
\newcommand{\bp}{{\bf p}}
\newcommand{\bq}{{\bf q}}
\newcommand{\br}{{\bf r}}
\newcommand{\bR}{{\bf R}}
\newcommand{\up}{\uparrow}
\newcommand{\down}{\downarrow}
\newcommand{\fns}{\footnotesize}
\newcommand{\ns}{\normalsize}
\newcommand{\cdag}{c^{\dagger}}

\title {A phenomenological investigation of the integral and the differential versions of the $Kimber$-$Martin$-$Ryskin$
  $unintegrated$ parton distribution functions using two different constraints and the  $MMHT2014$ $PDF$ }
\author{N. Olanj$^\dag$}\altaffiliation {Corresponding author, Email:   {n\_olanj@basu.ac.ir}, Tel:+98-81-38381601}
\author{M. Modarres$^\ddag$}
\affiliation{$^\dag$Physics Department, Faculty of  Science, Bu-Ali
Sina University, 65178, Hamedan, Iran} \affiliation{$^\ddag$Physics
Department, University of  Tehran, 1439955961, Tehran, Iran.}
\begin{abstract}
We  previously investigated the compatibility of the
$Kimber$-$Martin$-$Ryskin$ ($KMR$) $unintegrated$
 parton distribution functions ($UPDF$) with the experimental data on the proton (longitudinal) structure
  functions ($PSF$ ($PLSF$)).  Recently  $Golec-Biernat$ and $Stasto$ ($GBS$) demonstrated that the differential version of $KMR$ prescription and the implementations of  angular
 (strong)
 ordering ($AOC$ ($SOC$))  constraints, cause the negative-discontinuous $UPDF$ with the ordinary parton distribution
functions ($PDF$) as the input, which
 leads to a sizable effect on the calculation of $PSF$.
In the present work, we
  use the  new  $MMHT2014$-$LO$-$PDF$ as the input and focus on the  $UPDF$  behaviors as was raised by $GBS$.
  The  resulting $PSF$ and  $PLSF$  are compared with  the $MSTW2008$-$LO$-$PDF$ and $MRST99$-$PDF$ and the 2014 data given by the $ZEUS$ and $H1$
   collaborations. The calculated $PSF$ and  $PLSF$ based
  on the integral prescription of the $KMR$-$UPDF$ with
 the  $AOC$ and the ordinary $PDF$ as the input are reasonably consistent with the experimental data. Therefore, they
 are approximately
   independent to the  $PDF$ (no need to impose cutoff on the  $PDF$).
   At very small $x$ regions because of the excess
   of gluons in the  $MMHT2014$-$LO$-$PDF$ and $MSTW2008$-$LO$-$PDF$, an increase in $PSF$ and  $PLSF$ is achieved. Finally, according to the $GBS$ report the differential version by using the cutoff independent $PDF$ produces results far from the experimental data.
\end{abstract}
 \maketitle
\section{Introduction}
The parton distribution functions ($PDF$), $a(x,Q^2)$ = $ xq(x,Q^2)$
and $xg(x, Q^2)$, in which $x$ and $Q$ are the longitudinal momentum
fraction and the factorization or hard scale, respectively, are the
main phenomenological objects in the high energy collisions
computations of particle physics. These $PDF$ usually can be
extracted from the experimental data via the parametrization
procedures which are
 constrained by the sum rules and a few theoretical assumptions. These functions which usually called $integrated$
parton distributions, satisfy the standard
$Dokshitzer$-$Gribov$-$Lipatov$-$Altarelli$-$Parisi$ ($DGLAP$)
 evolution equations \cite{1a,1b,1c,1d}. The $DGLAP$ evolution equations  are derived by integrating over
the parton transverse momentum up to $k_t^2 = Q^{2}$. Thus the usual
$PDF$ are not the $k_t$-dependent  distributions.

On the other hand, there exist plenty of experimental data on the
various events, such as the exclusive and semi-inclusive processes
in the high energy collisions in the $LHC$, which indicate the
necessity for computation of the $k_t$-dependent parton distribution
functions. These functions are $unintegrated$ over $k_t$ and  are
  called the $unintegrated$ parton distribution functions ($UPDF$). The $UPDF$ are the two-scale dependent functions that
  can be generated via the $Ciafaloni$-$Catani$-$Fiorani$-$Marchesini$ ($CCFM$)
 formalism \cite{5,6,7,8}. Working in this framework is a hard and restrictive task. Also, there is not a complete quark version of
  the $CCFM $ formalism. Therefore, to overcome the complexity of the $CCFM$ equations and to calculate the $UPDF$,
    $Kimber$, $Martin$ and $Ryskin$ ($KMR$) \cite{91} proposed a procedure which is based on the standard $DGLAP$ equations
   in the $LO$ approximation, along with a modification due to the strong ordering condition ($SOC$) in transverse momentum of the real parton emission, which
comes from the coherence effect \cite{92}. The prescription along
with $SOC$  was further modified in the reference \cite{9} due to
the angular ordering condition ($AOC$), which is the key
   dynamical property of the $CCFM$ formalism (it is semi-$NLO$ formalism).

In our previous works \cite{13,14,14p}, to validate $KMR$ approach,
we have utilized the $unintegrated$ parton distribution functions
 in the $KMR$ $k_t $-factorization procedure by using the set of $MRST99$ \cite{MRST} and $MSTW2008$-$LO$ \cite{MSTW}
$PDF$ as the inputs to calculate the proton structure function and
the proton longitudinal structure function. Also, we successfully
used the $UPDF$ of the $KMR$ approach  to calculate the inclusive
production of the $W$ and $Z$ gauge
 vector bosons \cite{WZ,z}, the semi-$NLO$ production of Higgs bosons \cite{H} and the production of forward-center and
 forward-forward di-jets \cite{di}.

Recently  $Golec-Biernat$ and $Stasto$ ($GBS$) \cite{GBS} pointed
out that different
 versions of $KMR$ prescriptions as well as implementations of  angular
 ordering ($AOC$) and strong ordering ($SOC$) constraints, can cause negative and discontinuous $UPDF$ with the collinear global parton distribution functions ($PDF$) as the input that come from a global fit to data using the conventional
collinear approximation, which in turn
 especially can cause a sizable effect on the calculation of proton structure
 functions.  They showed that despite seemingly mathematical equivalence between
   the different
 versions of $KMR$ prescriptions with the same constraints, different results are obtained using the ordinary $PDF$ as the input (see the figure 1 of the reference \cite{GBS} ). Also, they have shown that the integral form $KMR$-$UPDF$ by using the ordinary $PDF$ and the cutoff dependent $PDF$ as inputs, gives approximately the same results (see the figure 4 of the reference \cite{GBS} ), in contrast to the differential form. They conclude that, this un-physical behavior happens in the differential form $KMR$
 prescription (see the equation (10) of $GBS$, the
 references \cite{9,Golec31} and the section $II$ of present
 report), otherwise one should impose cut off on the input $PDF$ . As it is stated in the reference \cite{watt2004}, the application of the integrated
 $PDF$ in the last evolution step should be generated through a new
 global fit to the data using the $k_t$-factorization procedures.
 This was estimated to lower the proton structure functions by 10
 per cent \cite{watt2004} (if one ignores this $k_t$-factorization
 fitting).

In the present work, following our previous investigations, we
intend to calculate the proton structure functions and the proton
longitudinal structure functions by using the different versions of
the $KMR$ $k_t $-factorization procedure \cite{9} and taking into
account the $PDF$ of $Martin$ et al. i.e., $MMHT2014$-$LO$ \cite{22}
as the input. The results of the integral version with $AOC$  are
compared with our previous studies based on the $MRST99$ and
$MSTW2008$-$LO$ input $PDF$ and the data given by the $ZEUS$
\cite{ZEUS} and $H1$  \cite{H1} collaborations. In general, it is
shown that our calculations are reasonably consistent with the
experimental data and, by a good approximation,
   they are independent of the input $PDF$. It is also shown that the calculated proton structure function and  the proton
   longitudinal structure function based
  on the integral prescription of the $KMR$-$UPDF$ with
 the  $AOC$ constraint and the ordinary $PDF$ as the input are reasonably consistent with the experimental data. Therefore, they
 are approximately
   independent to the  $PDF$  i.e. no need to impose cutoff on the  $PDF$.
   However, at very small $x$ regions because of the excess of gluons in the input $PDF$ of the $MMHT2014$-$LO$ and $MSTW2008$-$LO$, a better agreement is achieved (see the panels $Q^2$=$12$ $GeV^2$). Finally, according to the $GBS$ report by considering the
   integral prescription of the $KMR$-$UPDF$ (see the figure 1 of the reference
    \cite{GBS} and compare the solid curves of the left and right panels together)
    and the differential version of the $KMR$-$UPDF$, and using the cutoff independent $PDF$,
    we show the integral version with the $SOC$ constraint and the differential version produces results far from experimental data than the integral version with $AOC$ constraint especially as the hard scale is increased.

So the paper is organized as follows: in the section $II$ we give a
brief review of the different versions of the $KMR$ approach
\cite{9} for the extraction of the $UPDF$ form, regarding the
phenomenological $PDF$.
   The formulation of $F_{2}(x, Q^2)$ and $F_{L}(x, Q^2)$ based on the $k_t $-factorization approach are given in the section $III$.
   Finally, the section $IV$ is devoted to results, discussions, and conclusions.
\section{A brief review of the $KMR$ approach}
The $KMR$ \cite{9} approach was developed to calculate the $UPDF$,
 $f_a(x, k_t^2, Q^{2})$, by using the given $PDF$, ($a(x, Q^2)$ = $ xq(x, Q^2)$ and $xg(x, Q^2)$),
 and the corresponding splitting functions  $P_{aa^{\prime}}(x)$ at leading order ($LO$).
 This approach is the modification to the standard $DGLAP$ evolution equations by imposing
 the angular ordering constraint ($AOC$), which is the consequence of coherent gluon emissions
 (see below for the case of strong ordering constraint).
The $KMR$ approach has two different versions that have a seemingly
mathematical equivalence.
\\1. Integral form:
\\
In integral form of the $KMR$ approach the separation of the real
and virtual contributions in the $DGLAP$ evolution chain at the $LO$
level leads to the following forms for the quark and the gluon
$UPDF$:
\begin{eqnarray}
f_{q}(x,k_{t}^2,Q^{2})&=&T_q(k_t,Q)\frac{\alpha_s({k_t}^2)}{2\pi}
\nonumber\\&\times&
\int_x^{1-\Delta}dz\Bigg[P_{qq}(z)\frac{x}{z}\,q\left(\frac{x}{z},
{k_t}^2 \right)\cr &+& P_{qg}(z)\frac{x}{z}\,g\left(\frac{x}{z},
{k_t}^2 \right)\Bigg],
 \label{eq:8}
\end{eqnarray}
\begin{eqnarray}
f_{g}(x, k_{t}^2, Q^{2})&=&T_g(k_t, Q)\frac{\alpha_s({k_t}^2)}{2\pi}
\nonumber\\&\times& \int_x^{1-\Delta}dz\Bigg[\sum_q
P_{gq}(z)\frac{x}{z}\,q\left(\frac{x}{z} , {k_t}^2 \right) \cr &+&
P_{gg}(z)\frac{x}{z}\,g\left(\frac{x}{z} , {k_t}^2 \right)\Bigg],
 \label{eq:9}
\end{eqnarray}
respectively, while survival probability factor $T_a$ is evaluated
from:
\begin{eqnarray}
T_a(k_t,
Q)&=&\exp\Bigg[-\int_{k_t^2}^{Q^2}\frac{\alpha_s({k'_t}^2)}{2\pi}\frac{{dk'_t}^{2}}{{k'_t}^{2}}
 \sum_{a'}\int_0^{1-\Delta}dz'P_{a'a}(z')\Bigg].
 \label{eq:5}
\end{eqnarray}
In this approach only at the last step of the evolution does the
dependence on the second scale, $Q$, get introduced into the $UPDF$.
\\
2. Differential form:
\\ The differential form of the $KMR$ approach generates  $UPDF$  by using the derivation of the integrated $PDF$, as follows:
\begin{eqnarray}
f_{a}(x, k_{t}^2, Q^{2})&=&\frac{\partial}{\partial ln {\lambda}^2}
[ a(x, {\lambda}^2)T_a(\lambda, Q)]\Bigg\vert_{\lambda=k_t},
 \label{eq:differential}
\end{eqnarray}
where $T_a$ obtained from equation (\ref{eq:5}).
\\The required $PDF$ are provided as the input, using the libraries
$MRST99$ \cite{MRST}, $MSTW2008$ \cite{MSTW} and
 $MMHT2014$ \cite{22},
 where the calculation of the
single-scaled functions are carried out using the deep-inelastic
scattering ($DIS$) data on the $F_{2}(x, Q^2)$ structure function of
the proton. The cutoff, $\Delta=1-z_{max}=\frac{k_t}{Q+k_t}$ , is
determined by imposing the  $AOC$ on the last step of the
evolutionary, to prevent the $z=1$ singularities in the splitting
functions, which arise from the soft gluon emission. Also, $T_a(k_t,
Q)$ is considered to be unity for $k_t>Q$. This constraint and its
interpretation in terms of the angular ordering condition gives the
integral form of the $KMR$ approach a smooth behavior over the
small-$x$ region, which is generally governed by the
$Balitsky$-$Fadin$-$Kuraev$-$Lipatov$ ($BFKL$) evolution equation
\cite{23,24}. Notice that considering $T_a(k_t, Q)$=1 for $k_t>Q$,
the differential form of the $KMR$ approach is converted to the
following equation:
\begin{eqnarray}
f_{a}(x, k_{t}^2, Q^{2})&=&\frac{\partial}{\partial ln {\lambda}^2}
[ a(x, {\lambda}^2)]\Bigg\vert_{\lambda=k_t}.
 \label{eq:differential1}
\end{eqnarray}
As we stated above to prevent the $z=1$ singularities in the
splitting functions, which arise from the soft gluon emission, two
types of cutoffs, $\Delta$, were introduced, such that in the
equations (1), (2) and (3), $x$ to be less than $(1-\Delta)$:
\\
1. The strong ordering constraint ($SOC$) on the transverse momentum
of the real parton emission in the $DGLAP$ evolution: $\Delta =
\frac{k_t}{Q}$. In this case, the nonzero values of the $UDPF$ are
given for $k_t \leq Q(1-x)$  and therefore, we always have $k_t<Q$
and $T_a<1$.
\\
2. The angular ordering constraint ($AOC$) that we explained above,
which is the key dynamical property of the $CCFM$ formalism: $\Delta
=\frac{k_t}{Q+k_t}$, so the nonzero values of the $UDPF$ are given
for $k_t \leq Q(\frac{1}{x}-1)$ and  $T_a$ is considered to be unity
for $k_t> Q$ (see $GBS$).
\section{A glimpse of $F_{2}(x, Q^2)$ and $F_{L}(x, Q^2)$ in the $k_t $-factorization approach}
Here we briefly describe the different steps for calculations of the
proton structure functions
 ($F_{2}(x, Q^2)$) and the proton longitudinal structure functions ($F_{L}(x, Q^2)$) in
 the $k_t $-factorization approach. The $k_t $-factorization
 approach was discussed in several works, for example the references \cite{7,new1,new4,new5}.
 Since the gluons in the proton can only contribute to
  structure functions through the intermediate quark, so one should calculate the proton
  structure functions in the $k_t $-factorization
   approach by using the gluons and quarks $UPDF$. The $unintegrated$ gluons
and quarks contributions to $F_{2}(x, Q^2)$  and $F_{L}(x, Q^2)$
come from the subprocess
 $g \rightarrow q \overline{q}$ and $q \rightarrow qg$, respectively (see the figure 6 of the reference \cite{27}).
The relevant diagrams by considering a physical gauge for the gluon,
i.e., $A^{\mu}q^{\prime}_{\mu}=0 $ $(q^\prime=q+xp)$, are those
shown in the figure 1 (the figure 7 of the reference \cite{14}).
\subsection{The proton structure functions ($F_{2}(x,Q^2)$)}
The contributions for the diagrams shown in the figure 1 (the figure
7 of the reference \cite{14})  may be written in the $k_t
$-factorization form, by using the $unintegrated$ parton
distributions which are generated through the $KMR$ approach, as
follows for the gluons:
\begin{eqnarray}
F_2^{g \rightarrow q \overline{q}}(x, Q^2) &=&\sum_{q} e_q^2
\frac{Q^2}{4{\pi}^2} \int\frac{dk_t^2}{k_t^4}\int_0^{1}d\beta\int
d^2\kappa_t \alpha_s(\mu^2) f_g\left(\frac{x}{z}, k_t^2,
\mu^2\right)
 \Theta(1-\frac{x}{z})\nonumber\\
\Bigg \lbrace  [\beta^2 &+& (1-\beta^2)] (
\frac{\bf{\kappa_t}}{D_1}-\frac{(\bf{\kappa_t}-\bf{k_t})}{D_2} )^2 +
[m_q^2+4Q^2\beta^2(1-\beta)^2] (\frac{1}{D_1}-\frac{1}{D_2})^2 \Bigg
\rbrace
 \label{eq:2},
\end{eqnarray}
In the above equation, in which the graphical representations of
$k_t$ and $\kappa_t$ were introduced in the figure 1 (the figure 7
of the reference \cite{14}), the variable $\beta$ is defined as the
light-cone fraction of the photon momentum carried by the internal
quark \cite{9}. Also, the denominator factors are:
\begin{eqnarray}
D_1&=&\kappa_t^2+\beta(1-\beta)Q^2+m_q^2,\nonumber\\D_2&=&({\bf
{\kappa}_t} -{\bf k_t})^2 +\beta(1-\beta)Q^2+m_q^2
 \label{eq:3},
\end{eqnarray}
and
\begin{eqnarray}
\frac{1}{z}=1+\frac{\kappa_{t}^{2}+m_q^2}{(1-\beta)
Q^2}+\frac{k_t^2+\kappa_t^2-2{\bf \kappa_t}.{\bf k_t}+m_q^2}{\beta
Q^2},
 \label{eq:a}
\end{eqnarray}
As in the references \cite{9,28}, the scale $\mu$ which controls the
$unintegrated$ gluon and the $QCD$ coupling constant  $\alpha_s$ is
chosen as follows:
\begin{eqnarray}
\mu^2=k_t^2+\kappa_t^2+m_q^2.
 \label{eq:b}
\end{eqnarray}
For the charm quark, $m$ is taken to be $m_c = 1.27$ $GeV$, and $u$,
$d$ and $s$ quarks masses are neglected.
\\And for the quarks,
\begin{eqnarray}
{F_2}^{q\rightarrow qg}(x,Q^2)=\sum_{q=u,d,s,c} &e_q^2&
\int_{k_0^2}^{Q^2}
\frac{d\kappa_t^2}{\kappa_t^2}\frac{\alpha_s(\kappa_t^2)}{2\pi}\int_{k_0^2}^{\kappa_t^2}\frac{dk_t^2}{k_t^2}
\int_{x}^{1-\Delta}dz\nonumber\\ &\Bigg[& f_q \left(\frac{x}{z},
k_t^2, Q^2\right)+f_{\overline{q}} \left(\frac{x}{z}, k_t^2,
Q^2\right) \Bigg] P_{qq}(z)
 \label{eq:d}.
\end{eqnarray}
It should be noted that the above relations are true only for the
region of the perturbative $QCD$. The $unintegrated$ parton
distribution functions are not defined for $k_t<k_0$, i.e., the
$non$-perturbative region. So, according to the reference
\cite{new6}, $k_0$ is chosen to be about 1 $GeV$, which is around
the charm mass in the present calculation, as it should be.
Therefore, the contribution of the $non$-perturbative region for the
gluons   is  approximated \cite{9}, as follows:
\begin{eqnarray}
\int_{0}^{k_0^2} \frac{dk_t^2}{k_t^2} &f_g& (x, k_t^2, \mu^2)
\Bigg[\sum_{q} e_q^2 \frac{Q^2}{4{\pi}^2} \int_0^{1}d\beta\int
d^2\kappa_t \frac{\alpha_s(\mu^2)}{k_t^2} \Theta(1-\frac{x}{z})
 \nonumber\\ &\Bigg \lbrace &  [\beta^2 + (1-\beta^2)] (
\frac{\bf{\kappa_t}}{D_1}-\frac{(\bf{\kappa_t}-\bf{k_t})}{D_2} )^2 +
[m_q^2+4Q^2\beta^2(1-\beta)^2] (\frac{1}{D_1}-\frac{1}{D_2})^2 \Bigg
\rbrace\Bigg] \nonumber\\ &\simeq & xg(x, k_0^2)T_g(k_0,
\mu){\Bigg[\; \Bigg]}_{k_t=a} \label{eq:c},
\end{eqnarray}
where $a$ is a suitable value of $k_t$ between $0$ and $k_0$, which
its value is not important to the non-$perturbative$ contribution.
\\And for the quarks,
 \begin{eqnarray}
{F_2}^{q(non-perturbative)}(x, Q^2)= \sum_{q} e_q^2 ( xq(x,
k_0^2)+x\overline{q}(x, k_0^2) ) T_q(k_0, Q)
 \label{eq:f}.
\end{eqnarray}
Finally, the structure function $F_2(x, Q^2)$ is given by the sum of
the gluon contributions, the equations (\ref{eq:2}) and
(\ref{eq:c}), and the quark contributions, the equations
(\ref{eq:d}) and (\ref{eq:f}).
\subsection{The proton structure functions ($F_{L}(x, Q^2)$)}
In the equation (\ref{eq:olanj}) \cite{FL,new6,new7,new8}, i.e. the
formulation of $F_{L}(x, Q^2)$, the first term comes from the $k_t
$-factorization which explains the contribution of the $UPDF$ into
the $F_L$. This term is derived with the use of a pure gluon
contribution. However, it only counts the gluon contributions coming
from the perturbative region, i.e., for $k_t > 1$ $GeV$, and does
not have anything to do with the $non$-perturbative contributions.
Therefore, the third term is the gluon $non$-perturbative
contribution which can be derived from the $k_t $-factorization term
with the use of a variable-change, i.e., $y$, that carries the
$k_t$-dependent as follows:
 \begin{eqnarray}
y=x\bigg(1+\frac{{\kappa_t^{'}}^2+m_q^2}{\beta(1-\beta)Q^2}\bigg)
 \label{eq:f1},
\end{eqnarray}
while $\bf\kappa_t^{'}$  is defined as $\bf \kappa_t^{'}=\bf
\kappa_t- (1-\beta)\bf k_t$. Also, the second term is a calculable
quark contribution in the longitudinal structure function of the
proton, which comes from the collinear factorization:
\begin{eqnarray}
F_L (x, Q^2) =&\frac{Q^4}{\pi^2}&\sum_{q} e_q^2
 \int\frac{dk_t^2}{k_t^4} \Theta(k^2-k_0^2)\int_0^{1}d\beta\int
d^2\kappa_ t \alpha_s(\mu^2) \beta^2(1-\beta)^2\left
(\frac{1}{D_1}-\frac{1}{D_2}\right)^2\nonumber\\ &\times &
f_g\left(\frac{x}{z}, k_t^2, \mu^2\right) +\frac{\alpha_s(Q^2)}{\pi}
\frac{4}{3}  \int_x^{1}\frac{dy}{y} (\frac{x}{y})^2 F_2 (y,
Q^2)\nonumber\\ &+&\frac{\alpha_s(Q^2)}{\pi}\sum_{q} e_q^2
\int_x^{1}\frac{dy}{y}(\frac{x}{y})^2 (1-\frac{x}{y}) y g(y, k_0^2)
\label{eq:olanj},
\end{eqnarray}
where the second term is (see \cite{FL}):
\begin{eqnarray}
\sum_{q} e_i^2 \frac{\alpha_s(Q^2)}{\pi} \frac{4}{3}
\int_x^{1}\frac{dy}{y}(\frac{x}{y})^2 [q_i(y, Q^2)+\overline{q}_i(y,
Q^2)]\label{eq:nn},
\end{eqnarray}
while the variables of the above equation are the same as those
expressed in relation to the  proton  structure function $(F_2(x,
Q^2))$.
\section{Results, discussions and conclusions}
As it was described in the section $II$, the $KMR$ approach was
developed to calculate the $UPDF$, by using the given the global
fitted $PDF$ as the input. To make the comparison more clear, the
typical inputs, the gluon and the up quark $PDF$ considering the
$PDF$ uncertainties at scale $Q^2$ = $60$ $GeV^2$, by using the
$MRST99$ \cite{MRST} , $MSTW2008$-$LO$ \cite{MSTW}  and
$MMHT2014$-$LO$ \cite{22}, are plotted in the figure 2.

The behavior of these integrated $PDF$ were discussed in detail in
the related  references \cite{MRST,MSTW,22}. The $MMHT2014$ $PDF$
supersede the $MSTW2008$ parton sets and these $MSTW2008$ $PDF$
supersede the previously available $MRST$ sets. Also, as shown in
the figure 2, these three sets are different at the very low $x$
region, that is the region where the transverse momentum becomes
important. Especially for the gluons, the $MRST$ parton sets are
very different from the other collaborations. Given the above
mentioned issues, to study the effect of increasing the contribution
of the gluon and the process of evolution in the $MRST$ set, we were
motivated to consider all of these three sets of $PDF$ in our
calculations. They are different (especially for the gluons $PDF$)
at very low $x$ regions (this is the region where the transverse
momentum becomes important) and they look similar at the large $x$
regions.

Respectively, in the figures 3 and 4, the proton structure functions
($F_{2}(x, Q^2)$) and the proton longitudinal structure functions
($F_{L}(x, Q^2)$) in the framework of the integral form of the $KMR$
approach with the application of the $AOC$ constraint, by using
central values of the $MRST99$, the $MSTW2008$- $LO$ and
$MMHT2014$-$LO$ $PDF$ inputs, versus $x$, for $Q^2$ = $12$, $60$,
$120$ and $250$ $GeV^2$ are plotted. Then, the predictions of this
approach for the proton structure functions ($F_{2}(x, Q^2)$) and
the proton longitudinal structure functions ($F_{L}(x, Q^2)$) are
compared to the recent measurements of $ZEUS$  \cite{ZEUS} and $H1$
\cite{H1}  experimental data.

The results emphasize that (as it was shown in the references
\cite{14,15,16,17,18,19}), the $KMR$ approach suppresses the
discrepancies between the inputs $PDF$, in which the
 presence of cutoff $AOC$ ($\Delta=\frac{k_t}{Q+k_t}$) has the key role.
 This property leads the outputs $UPDF$ which are more similar.
  As a result,
the $UPDF$ generated via applying three different inputs $PDF$ have
less discrepancies and in turn, each sets of $F_{2}(x, Q^2)$ or
$F_{L}(x, Q^2)$ values with above $PDF$ are very close to each
other. Although, in all of the panels  of the figures 3 and 4, the
discrepancies grow up with reduction of $x$ but it happens at very
lower rate than the $PDF$ themselves (see the figures 2).

It should be noted that the results of using the $MMHT2014$-$LO$
$PDF$ and the $MSTW2008$- $LO$ $PDF$ inputs at very low $x$ regions
are closer to the experimental data than the inputs of the $MRST99$.
This indicates that inclusion of more gluons in the very small $x$
region is important (see panels $Q^2$ = 12 $GeV^2$ in the figures 3
and 4).

In the different panels of the figure 5, similar to
\cite{18,19,GBS}, (note that in reference \cite{GBS}
 $xf_g(x, k_t^2, Q^2)/k_t^2$ is plotted), we plot the $UPDF$ (for the gluon and the up quark) with the input $MMHT2014$-$LO$ $PDF$ as a function of $k_t^2$ ($GeV^2$) for the two types of constraint discussed in the section $II$, i.e. $AOC$ and $SOC$, using the differential and integral forms
  of the $KMR$ approach. The hard scale is $Q^2 = 100$ $GeV^2$ and $x$ = 0.1, 0.01 and
0.001.

Despite seemingly mathematical equivalence between
   the differential and integral forms of $KMR$ prescription with the same constraints, the differences between them are manifested for the smaller
    $x$ values at the smaller transverse momentums (see that in gluon panels, the $UPDF$ of the two different versions with $AOC$
    constraint separated from each other
    at $k_t^ 2 \simeq {2,7}$  and 30 $GeV^2$ for the $x$ = 0.001, 0.01 and 0.1, respectively).  As $GBS$ reported, this difference is due to the fact that we used the usual global
    fitted $PDF$ instead of the cutoff dependent $PDF$ for generating the $UPDF$. As we expect from the relation of $x$ and
$\Delta$ discussed in the section $II$, the $SOC$ integral $UPDF$
become zero, when the transverse momentums become equal to the hard
scale while
 those of $AOC$ smoothly go to zero for large transverse momentum.

  But despite our expectation, the $SOC$ differential $UPDF$ with the
  global fitted $PDF$ as the input are nonzero for $ k_t>Q$.
      Because in this region, as discussed in the Section $II$, $T_a(k_t, Q)$ is considered to be unity, and the differential form of $KMR$ prescription (equation \ref{eq:differential} )
       turns into the equation (\ref{eq:differential1}) which is independent of  the cutoff for the global fitted $PDF$ as the input. As a result, as shown in the various panels
        in the figure 5, the differential $UPDF$ with $SOC$ and $AOC$ for $k_t>Q$ are the same and at the very large transverse momentums becomes larger
         than the $AOC$ integral $UPDF$ (see panels $x$=0.01, 0.001). Also, as $GBS$ reported, the differential version of $KMR$ prescription
          with the different constraints with the  the usual global
    fitted $PDF$ as the input leads to some un-physical results for large transverse momenta values.
          They are negative at  $k_t>Q$ for panels $x$=0.1
          and discontinuous at $k_t=Q$, that is a result of the discontinuity of the first derivative of the Sudakov form factor at $k_t=Q$.

         But, the curves obtained from the integral form for both constraints behave in a smooth way without any un-physical results.
          Therefore, as we pointed out above, and that the integral form
$KMR$-$UPDF$ by using the ordinary $PDF$ and the cutoff dependent
$PDF$ as inputs, gives approximately the same results (as the $GBS$
report), if we intend to use the usual global fitted $PDF$ as the
input for generating the $UPDF$, we can use only the integral
version of $KMR$ prescription.

The proton structure function ($F_{2}(x, Q^2)$) and the  proton
longitudinal structure functions ($F_{L}(x, Q^2)$) by using  the
integral and differential versions of the $KMR$ $k_t $-factorization
procedure for the $AOC$ and $SOC$ cutoffs are plotted in the figures
6 and 7 at hard scale 12, 60, 120 and 250 $GeV^2$, respectively. The
$F_{2}(x, Q^2)$ of the $LO$ collinear procedure and the experimental
data of $H1$ and $ZEUS$ are also given for comparison.

As the energy scale increase the difference between the integral
forms with the $AOC$ and $SOC$ cutoffs become more and those are
separated from each other specially at small $x$ values and the
$SOC$ results are below those of $AOC$. As far as present data are
concerned, the $AOC$ results are much more closer to the data with
respect to the $SOC$ cases. Regarding that the differential $UPDF$
with $SOC$ and $AOC$ for $k_t>Q$ are the same
 and at the very large transverse momentums becomes larger than the $AOC$ integral $UPDF$, the calculated proton structure functions and the proton longitudinal
  structure functions based on the $UPDF$ of the differential $KMR$ approach with $SOC$ and $AOC$ are the same by a good approximation and larger than those
   based on the $UPDF$ of the integral $KMR$ approach with $AOC$ at very small $x$ regions. Interestingly, despite some
    un-physical results for the differential form by using the usual global fitted $PDF$ as the input, approximately, the proton structure
   functions and the proton longitudinal structure functions based on the differential $UPDF$ are consistent with the experimental data.
  By comparing the curves of the figure 6, it turns out that integral form of $KMR$ prescription with $AOC$ is
   more consistent with the experimental data and the pure $LO$ collinear procedure than the others. Therefore, our structure function
   calculations  in the framework of the integral form of the $KMR$ approach for the $AOC$ constraint  confirm the conclusion which was made by $GBS$ that it is possible to use the usual global fitted $PDF$ instead of the cutoff dependent $PDF$ for generating the $UPDF$ of the $KMR$ approach by a good approximation.

 In conclusion, it was shown that calculated proton structure
functions and the proton longitudinal structure functions based on
the $UPDF$ of the integral version of the $KMR$ approach for the
$AOC$ constraint are reasonably consistent with the experimental
data and, by a good approximation, they are independent to the input
$PDF$. Therefore, they can be widely used in
 the calculations related to the particle physics phenomenology \cite{m1}.
 On the other hand, even the $k_t $-factorization and the $KMR$ approach can
 explain the shadowing effect in nuclei better than other nuclear physics indications \cite{m2,m3}.
 On the other hand, different constraints cutoffs were investigated
 using the the integral and the differential formulations of the $KMR$ prescription. The
 results confirm the statement made by the $GBS$ that:  (1) According to the compatibility of the proton structure functions generated using $AOC$ integral $UPDF$ with the ordinary $PDF$ (the usual global fitted $PDF$) as the input, with the experimental data, it can be concluded that it is possible to use the usual global fitted $PDF$ instead of the cutoff dependent $PDF$ for generating the
 $UPDF$, especially because to fit the $PDF$ through the $UPDF$ is
 the cumbersome task. (2) As we pointed out above, due to some un-physical results for the differential form by using the ordinary $PDF$ as the input, as far as one
 used the integral form of the $KMR$ approach and the $AOC$ by using the ordinary $PDF$ as the input, there would not be any problem for the calculations of structure functions
 and hadron-hadron cross section in the framework of the $k_t$-factorization.
\begin{acknowledgements}
NO would like to acknowledge the University of Bu-Ali Sina for their
support. MM would also like to acknowledge the Research Council of
the University of Tehran for the grants provided for him.
\end{acknowledgements}

\newpage
 
\begin{figure}[h!]
\includegraphics[scale=0.6]{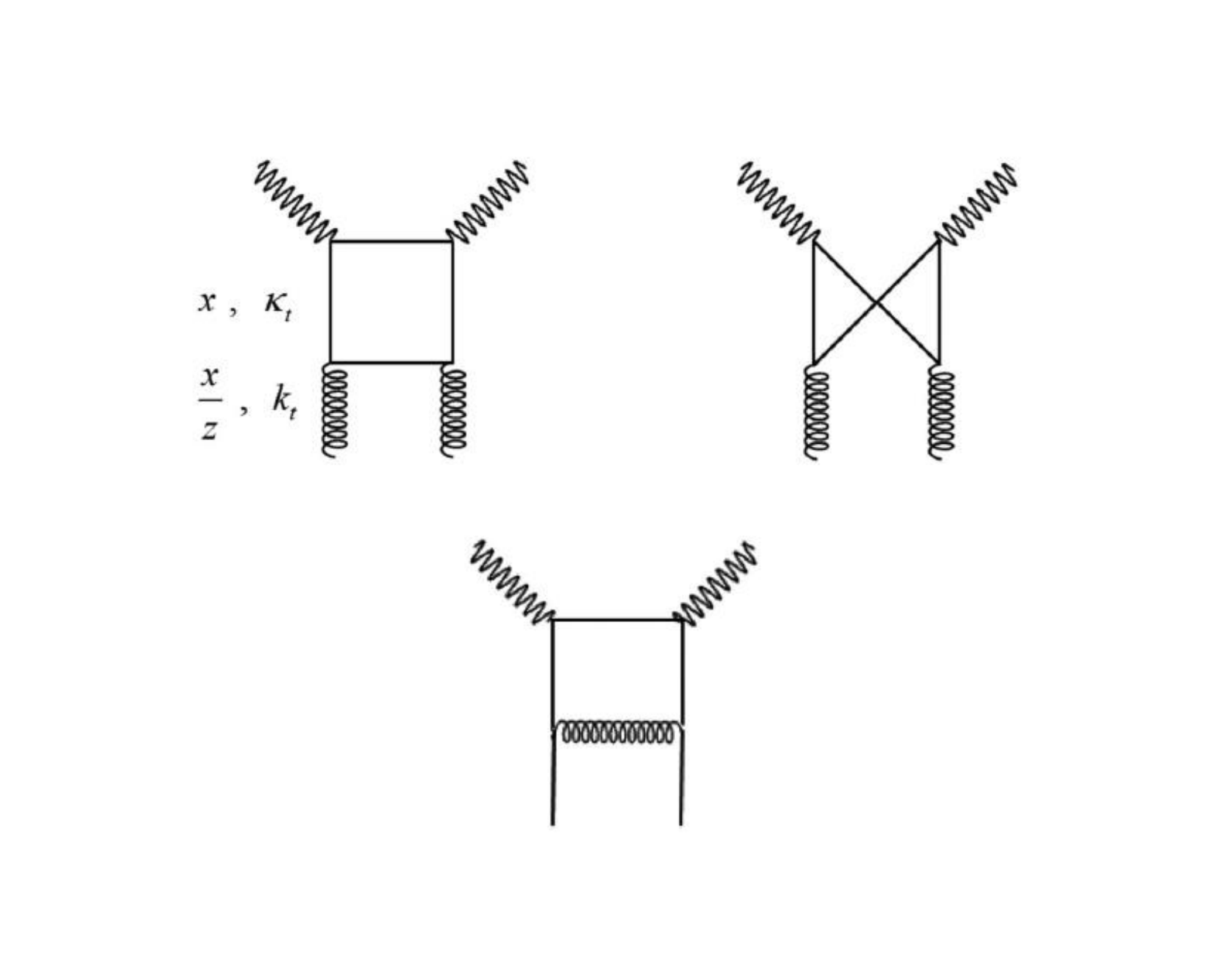}
\caption{The diagrams contributing in the calculation of the
structure functions $F_2(x,Q^2)$, which comes from the $g
\rightarrow q \overline{q}$ and $q\rightarrow qg$.}
\label{fig:1}
\end{figure}
\begin{figure}[h!]
\includegraphics[scale=0.6]{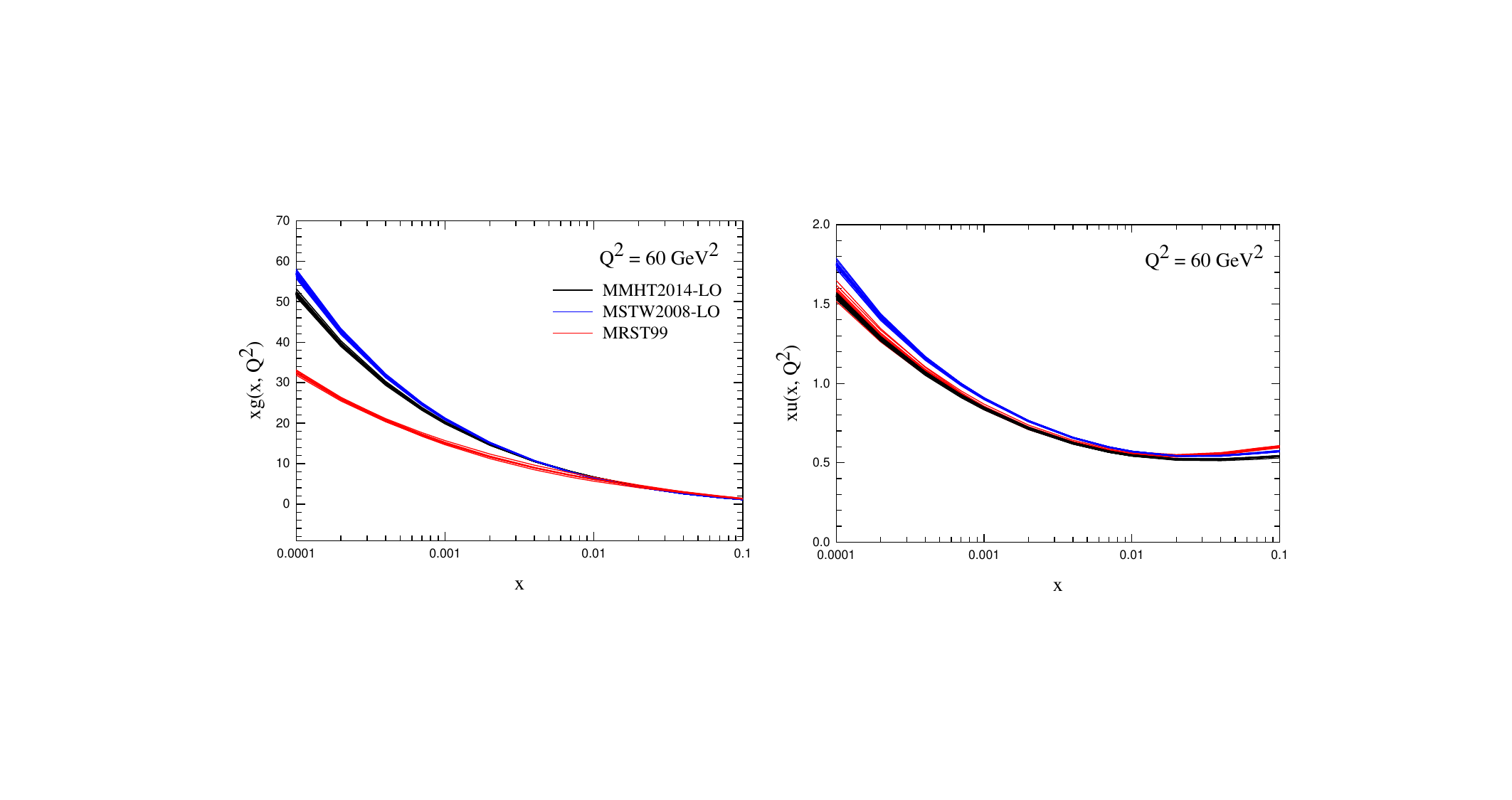}
\caption{The integrated gluon and up quark distribution functions
(see the text for detail).}\label{fig:2}
\end{figure}
\begin{figure}[h!]
\includegraphics[scale=0.6]{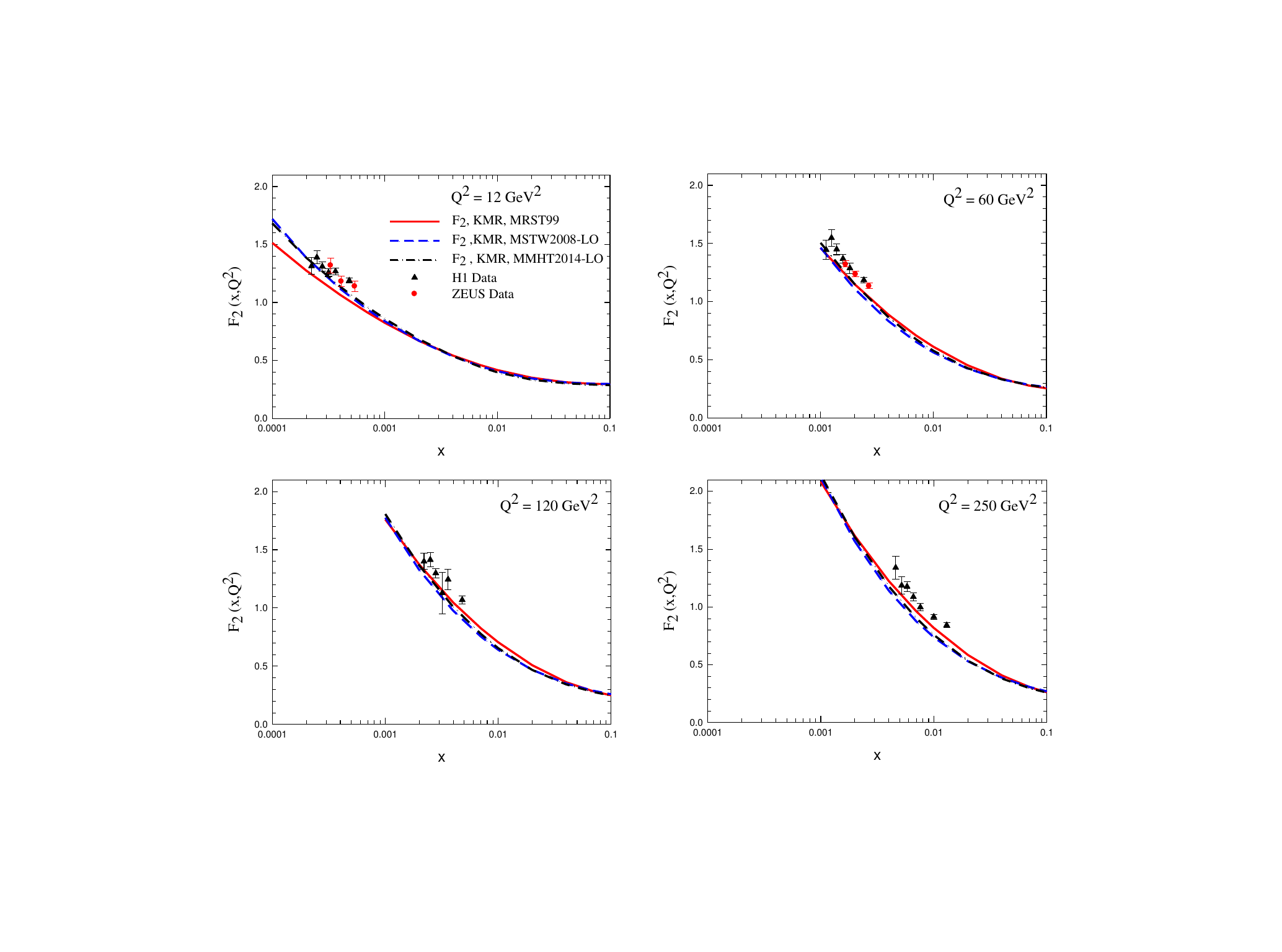}
\caption{The proton structure functions $F_{2}(x, Q^2)$ based
  on the integral form of the $KMR$ approach
 with the $AOC$ constraint as a function of $x$ for various $Q^2$ values, by using the $MRST99$ \cite{MRST}, the $MSTW2008$-
$LO$ \cite{MSTW} and the $MMHT2014$-$LO$ \cite{22} as the inputs,
are compared with the $ZEUS$ \cite{ZEUS} and $H1$ \cite{H1}
experimental data.}\label{fig:3}
\end{figure}
\begin{figure}[h!]
\includegraphics[scale=0.6]{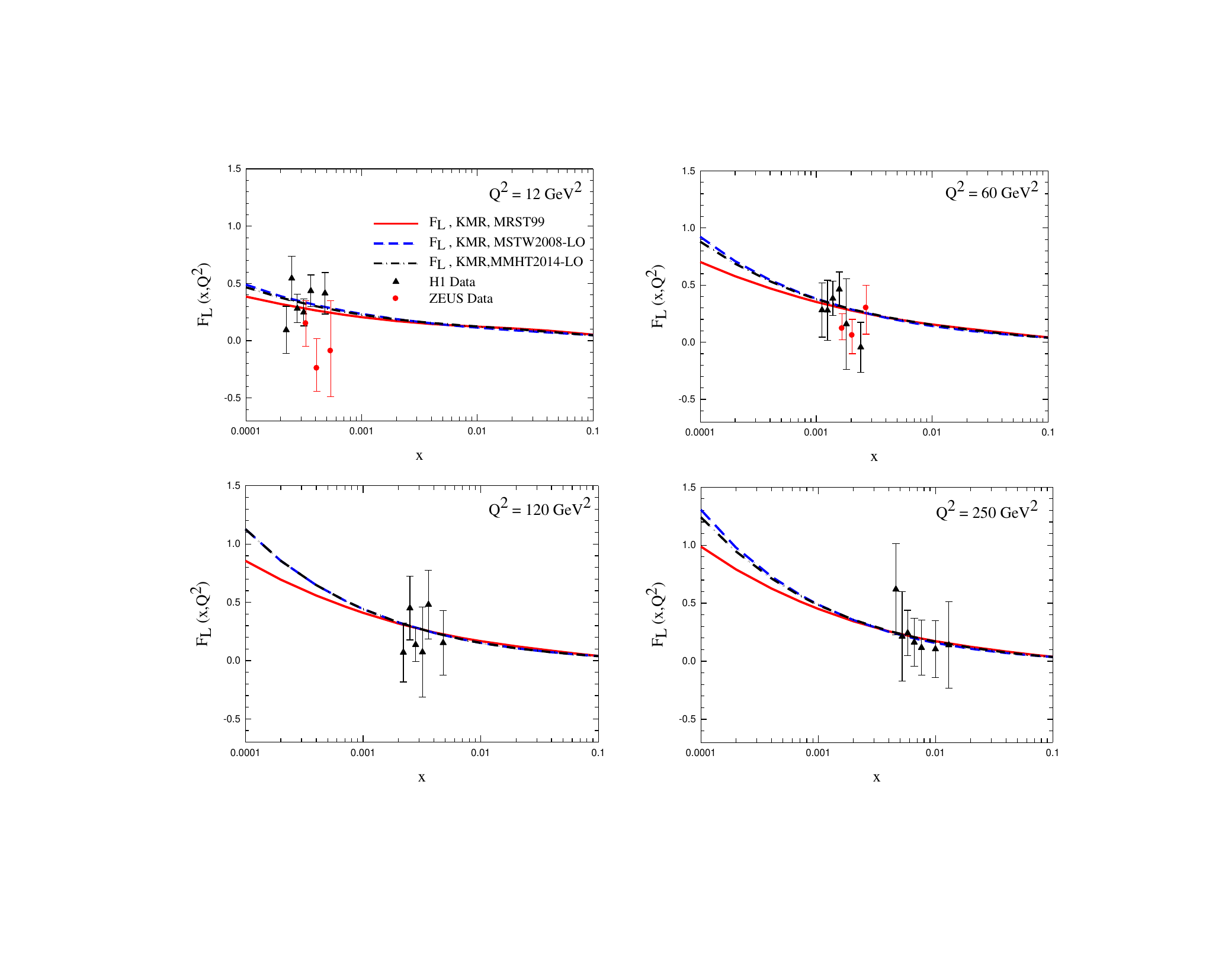}
\caption{The  proton longitudinal structure functions $F_{L}(x,
Q^2)$ based
  on the integral form of the $KMR$ approach
 with the $AOC$ constraint as a function of $x$ for various $Q^2$ values, by using the $MRST99$ \cite{MRST}, the $MSTW2008$-$LO$ \cite{MSTW} and the $MMHT2014$-$LO$ \cite{22} as the inputs,
are compared with the $ZEUS$ \cite{ZEUS} and $H1$ \cite{H1}
experimental data.}\label{fig:4}
\end{figure}
\begin{figure}[h!]
\includegraphics[scale=0.6]{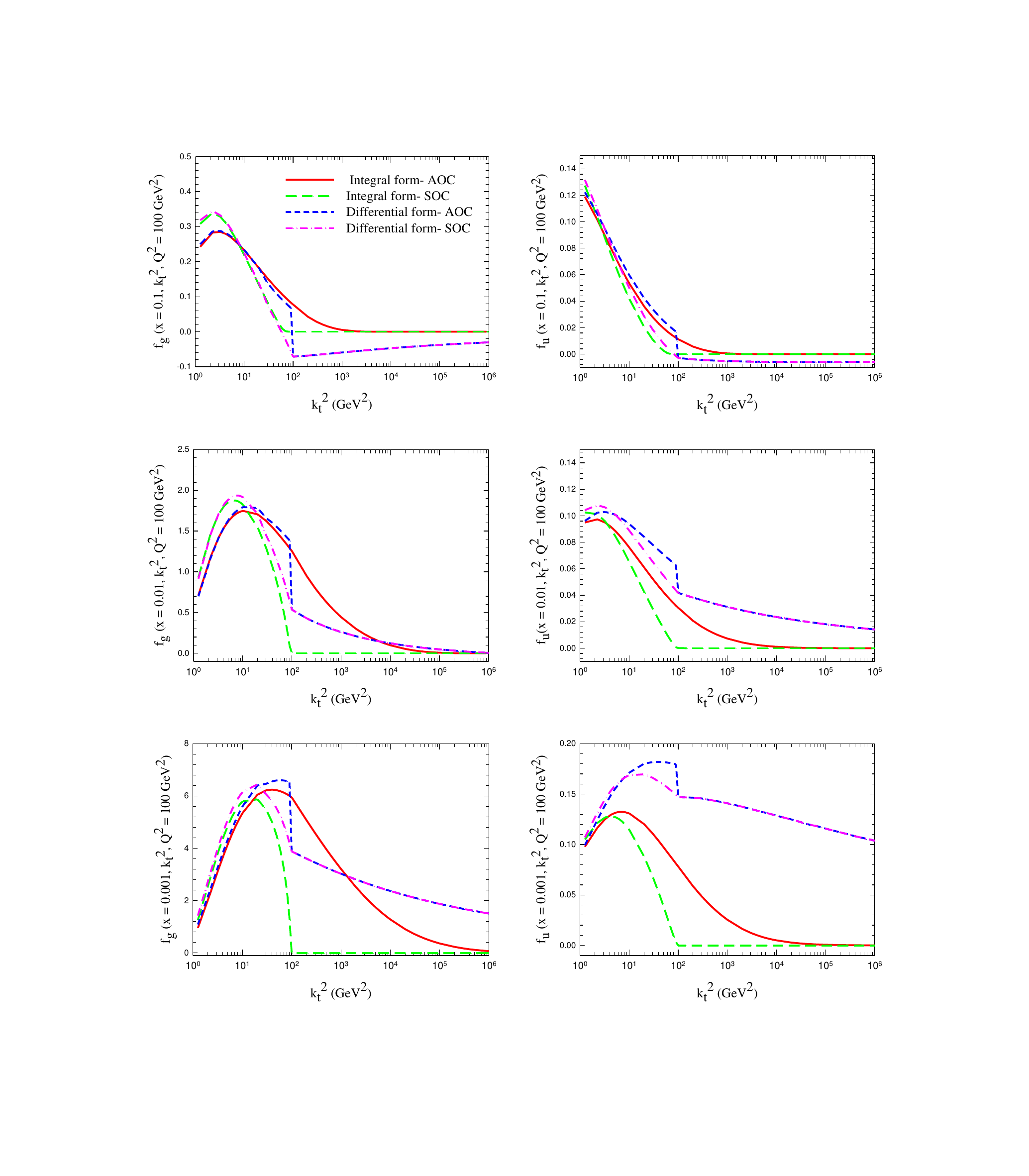}
\caption{The gluon and up quark  $UPDF$ as a function of $k_t^2$
($GeV^2$) for the different versions of the $KMR$ approach with the
$AOC$  and $SOC$ constraints and $x$ = 0.1, 0.01 and 0.001.
}\label{fig:5}
\end{figure}
\begin{figure}[h!]
\includegraphics[scale=0.6]{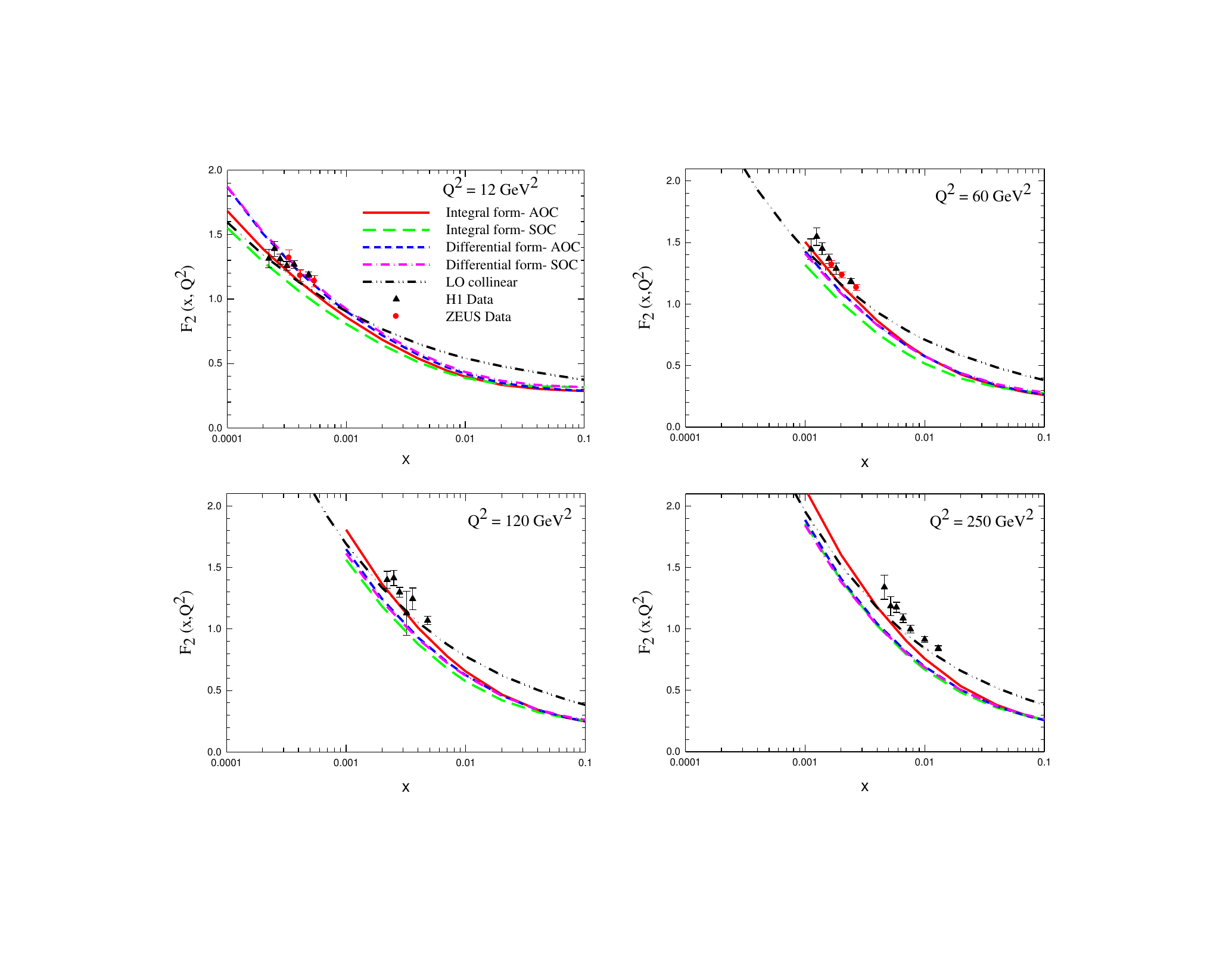}
\caption{The proton structure functions $F_{2}(x, Q^2)$ based on the
different versions of the $KMR$ approach as a function of $x$ for
various $Q^2$ values, by using  the $MMHT2014$-$LO$ \cite{22} $PDF$
as the
 inputs and the $AOC$  and $SOC$ constraints in
comparison with the $F_{2}(x, Q^2)$ of the $LO$ collinear procedure
with the input $MMHT2014$-$LO$ $PDF$ and the $ZEUS$ \cite{ZEUS} and
$H1$ \cite{H1} experimental data.}\label{fig:6}
\end{figure}
\begin{figure}[h!]
\includegraphics[scale=0.6]{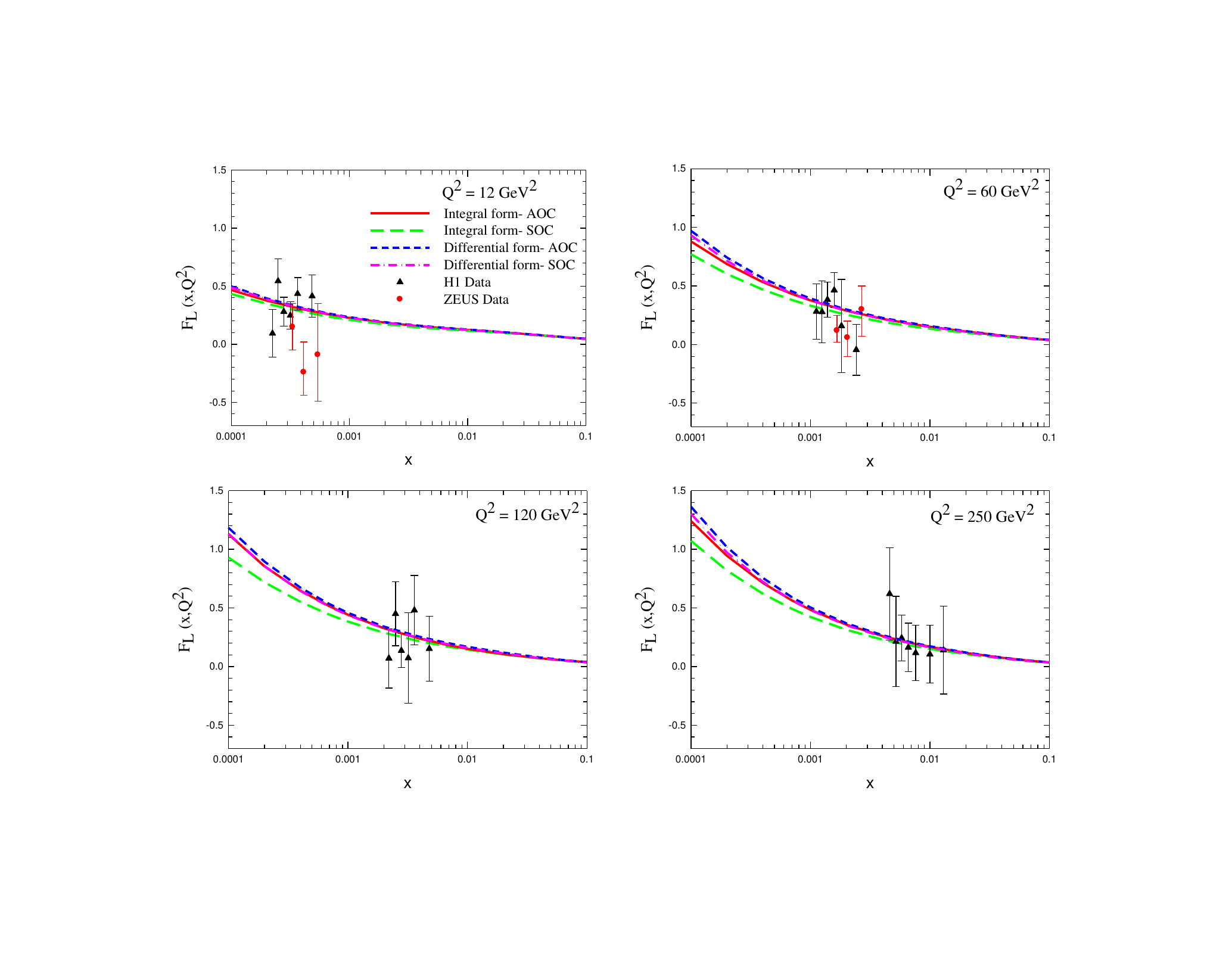}
\caption{The proton longitudinal structure functions $F_{L}(x, Q^2)$
based on the different versions of the $KMR$ approach
 as a function of $x$ for various $Q^2$ values, by using  the $MMHT2014$-$LO$ \cite{22} $PDF$ as the
 inputs and the $AOC$  and $SOC$ constraints in
comparison with the $ZEUS$ \cite{ZEUS} and $H1$ \cite{H1}
experimental data.}\label{fig:7}
\end{figure}
\end{document}